\documentclass[a4paper,11pt]{article}
\pdfoutput=1 

\usepackage{jcappub} 
\usepackage[uncertainty-mode=separate,list-units=single]{siunitx}
\usepackage{soul,xcolor}
\usepackage[T1]{fontenc} 
\usepackage[shortcuts]{extdash}
\usepackage[normalem]{ulem}
\usepackage[nameinlink,noabbrev]{cleveref}
\usepackage{calc}
\usepackage{amsmath}
\usepackage{bm}
\usepackage{amssymb}
\usepackage{siunitx}
\usepackage{enumitem}
\usepackage{subcaption}
\usepackage{changes}
\usepackage{appendix}
\usepackage[numbers]{natbib}
\usepackage{cancel}
\crefname{equation}{eq.}{eqs.}
\Crefname{Equation}{Eq.}{Eqs.}

\DeclareSIUnit\steradian{steradian}

\providecommand{\sorthelp}[1]{}

\begin{document}

\title{Reassessment of the dipole in the distribution of quasars on the sky}

\author[1,2]{Arefe Abghari,\note{Corresponding author.}}
\author[3]{Emory F. Bunn,}
\author[2]{Lukas T. Hergt,}
\author[2]{Boris Li,}
\author[2]{Douglas Scott,}
\author[2]{Raelyn M. Sullivan,}
\author[2]{Dingchen Wei}
\affiliation[2]{University of British Columbia, Vancouver, BC V6T1Z1, Canada}
\affiliation[3]{University of Richmond, Richmond, VA 23173, U.S.A.}

\emailAdd{arefeabghari@phas.ubc.ca}
\emailAdd{ebunn@richmond.edu}
\emailAdd{rsullivan@phas.ubc.ca}
\emailAdd{lthergt@phas.ubc.ca}
\emailAdd{dscott@phas.ubc.ca}

\abstract
{We investigate recent claims by Secrest et al. of an anomalously large amplitude of the dipole in the distribution of CatWISE-selected quasars on the sky.  Two main issues indicate that the systematic uncertainties in the derived quasar-density dipole are underestimated.  Firstly, the spatial distribution of the quasars is not a pure dipole, possessing low-order multipoles of comparable size to the dipole. These multipoles are unexpected and presumably caused by unknown systematic effects; we cannot be confident that the dipole amplitude is not also affected by the same systematics until the origin of these fluctuations is understood.  Secondly, the 50 percent sky cut associated with the quasar catalogue strongly couples the multipoles, meaning that the power estimate at $\ell=1$ contains significant contributions from $\ell>1$.  In particular, the dominant quadrupole mode in the Galactic mask strongly couples the dipole with the octupole, leading to a large uncertainty in the dipole amplitude.  Together these issues mean that the dipole in the quasar catalogue has an uncertainty large enough that consistency with the cosmic microwave background (CMB) dipole cannot be ruled out.  More generally, current data sets are insufficiently clean to robustly measure the quasar dipole and future studies will require samples that are larger (preferably covering more of the sky) and free of systematic effects to make strong claims regarding their consistency with the CMB dipole.}

\maketitle

\section{Introduction}
\label{sec:intro}
\begin{figure}
    \centering
    \includegraphics[width=1\linewidth]{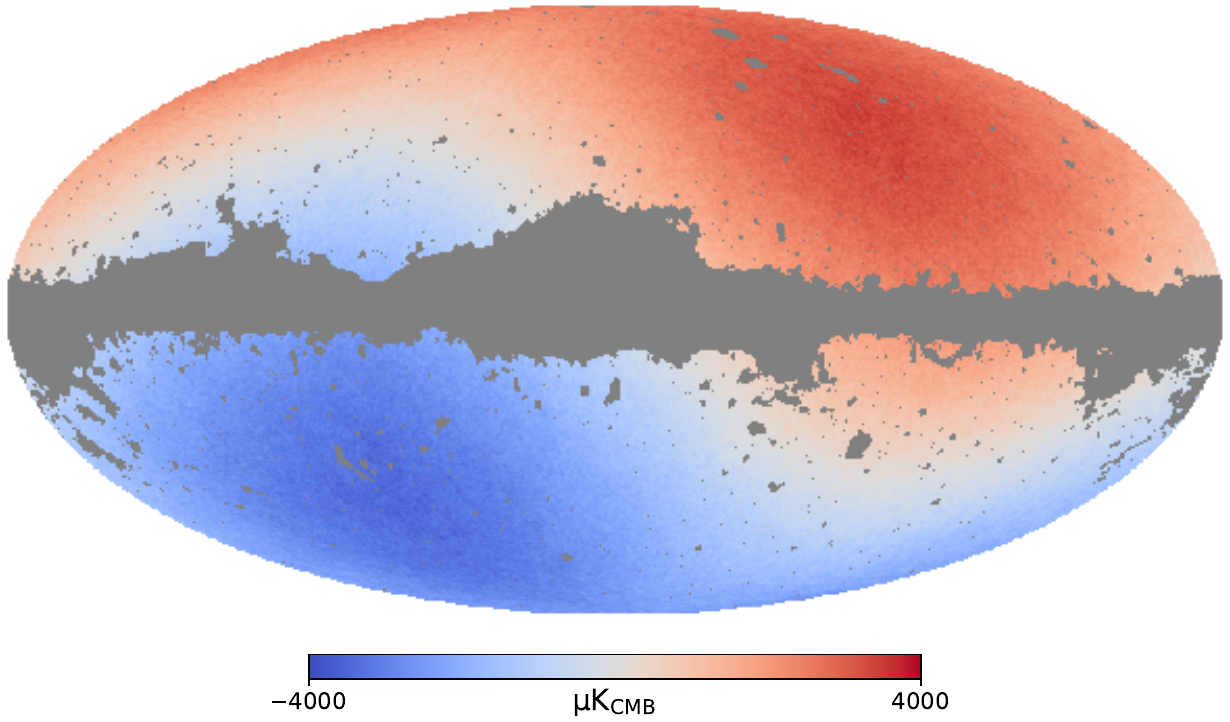}
    \caption{All-sky CMB map from \textit{Planck} with the monopole subtracted.  We specifically show the Public Release 4 data \citep{planck2020-LVII}, employing the \texttt{Commander} component-separation method.  A mask covering \SI{22}{\percent} of the sky has been applied ($f_\mathrm{sky}=\SI{78}{\percent}$), which removes much less of the sky than is needed for the quasar map analysis.  Even though underlying anisotropies and Galactic contamination have not been removed here, the dipole signal is much stronger than higher-order multipoles and dominates the image.
    }
    \label{fig:CMBDipole}
\end{figure}

The cosmological principle is the assumption that, on large enough scales, the Universe is homogeneous and isotropic. One consequence of this principle is that there should be no preferred direction or location in the Universe. The cosmological principle imposes symmetries that greatly simplifies to the Friedmann--Lema{\^\i}tre--Robertson--Walker~(FLRW) metric (for a historical discussion see, e.g., Section~2 of Peebles' book~\cite{Peebles1993}). According to this picture, the dipole of the cosmic microwave background (CMB) is solely caused by the motion of the Earth relative to the `rest frame' of the CMB, in which the sky is expected to be statistically isotropic. The temperature gradient caused by this movement is expressed as $\Delta T/T=(v/c)\cos\theta$~\cite{Peebles1968}, where $\theta$ is the angle from the direction of the motion. The most recent measurement of the speed of the Solar System with respect to the CMB rest frame comes from the \textit{Planck} satellite and is $v=\SI{369.82\pm0.11}{\kilo\meter\per\second}$ in the direction $l = \ang{264.021}\pm\ang{0.011}, b=\ang{48.253}\pm\ang{0.005}$ in Galactic coordinates \citep{planck2016-l01,planck2016-l03}.  \Cref{fig:CMBDipole} shows a map from  \textit{Planck}  that contains the Solar dipole; i.e.,\ the strong $\ell=1$ mode is very obvious. This is what a high signal-to-noise dipole looks like when it dominates over other multipoles and when the map does not require a large Galactic mask; this is in contrast to the quasar dipole that we discuss in the rest of the paper.

In addition to measuring the $\ell=1$ multipole of the CMB, there are several other ways to measure our cosmic motion, including the following: aberration and anisotropy modulation effects in the CMB \citep[e.g.][]{planck2013-pipaberration, planck2020-LVI, Saha2021}; determining the dipole in the peculiar velocity field of distant objects \citep{MRR2000,planck2013-XIII}; or summing the effect of large-scale structure on our local acceleration \citep[e.g.,][]{Erdogdu2006}.  If any of these tests of our motion gave results that did not match the CMB dipole velocity, then that would point to the existence of a large-scale non-adiabatic mode \citep[][]{Turner1991, Turner1992,ZibinScott2008} or some more fundamental breakdown of physics near the Hubble scale.  This motivates work comparing different observables that depend on our velocity.

Another way to measure our cosmological motion is to use the anisotropy of object counts or brightness using distant cosmological objects such as galaxies, quasars, or radio sources. This was first discussed by Ellis and Baldwin \citep{EllisBaldwin1984}. The dipole anisotropy of object counts across the sky is predicted to be (to order~$v/c$) 
\begin{equation}
    \delta N /N =[2 + x(1 + \alpha)](v/c) \cos(\theta) \equiv D\cos(\theta),
    \label{eq:qdipole}
\end{equation}
where $\alpha$ is the spectral index of the typical spectrum of an object (assuming that the sources have power-law spectra $S \propto \nu^{-\alpha}$), and $x$ is the slope of the cumulative number count as a function of limiting flux density, $N(>S)\propto S^{-x}$.  The amplitude of the dipole in this equation is denoted by~$D$.

There is a long history of attempts to measure the dipole of distant sources at a range of wavelengths \citep{MeiksinDavis1986, Lahav1987, Scharf2000, Watkins2023,Tiwari2024}. Among the studies done using radio or quasar catalogues, some have found that the dipole amplitude and direction align with the CMB dipole \citep{BlakeWall2002, Crawford2009, Gibelyou2012, Rubart2014, TiwariNusser2016, Darling_2022, Wagenveld2023, Wagenveld2024}. On the other hand, some studies have indicated a rough agreement with the CMB dipole direction, but have reported a higher than expected dipole amplitude \citep{Singal_2011, Rubart_2013, Colin2017, Tiwari_2019, Singal2019, Siewert2020, Oayda2024}. A common theme in all these studies has been the difficulty in controlling systematic effects; thus, various analysis approaches have been used in these papers. 

In a recent study, Secrest et al.~\cite{SecrestEtAl2021,SecrestEtAl2022} used the \textsc{CatWISE} catalogue \citep{catWISE2020} from the Wide-Field Infrared Survey Explorer (WISE)~\cite{WISE2010} to estimate the dipole in the distribution of quasars. They found that the amplitude of the quasar dipole appeared to be significantly larger than that expected from the CMB dipole, with a difference estimated to be at the $4.9\,\sigma$ level (for a normal distribution, one-sided). This has led to speculation in the literature of an unexpected contribution to the quasar dipole from large-scale structure~\cite{Das2021, Watkins2023,Whitford2023}; or for an intrinsic contribution to the CMB dipole \citep{KingEllis1973, Matzner1980,Ebrahimian2024}; or even more radically that the cosmological principle might be violated \citep{Allahyari2023, 
 Krishnan2022, Constantin2023}. 
 
Several studies have explored potential sources of error in dipole estimations \citep{Cheng2023}. As an example, Dalang et al.\ \citep{Dalang2021} and Guandalin et al.~\citep{Guandalin2023} specifically examined the influence of spectral index variations with redshift on the estimate of the quasar dipole.
In this paper, we investigate other sources of bias and uncertainty in the quasar dipole estimation. Specifically, we find (in \cref{sec:WISEquasars}) that the quasar density in the sky appears to be affected by systematic effects that cause non-uniformities or anisotropies on large angular scales.  The cause of these anisotropies is unknown, but since they cause fluctuations on large angular scales, one must assume that they affect the dipole, and hence comparison of the source dipole with the CMB dipole is not straightforward. Through a simple data split, we demonstrate that the data are most likely contaminated by non-quasar objects. In addition, we find (in \cref{sec:multipoleleakage}) that the applied masking significantly impacts the dipole estimator used by Secrest et al~\cite{SecrestEtAl2021}. Since higher-order multipoles leak into the lower-order multipoles when the map is strongly masked, this can dramatically affect the accuracy of the dipole moment estimation. We calculate the impact of higher multipoles on the dipole and assess the bias that this can introduce to this particular dipole estimator. We demonstrate that 
with these systematic effects in the data, the $p$-value associated with the measurement of the aforementioned dipole magnitude will be notably reduced, approximately to the $2\,\sigma$ significance level. We find (in \cref{sec:sim_real_qm}) that failure to account for masking in the analysis could lead to a biased estimation of the dipole amplitude and direction, and certainly an underestimate of the uncertainties. We attempt to estimate the quasar map dipole and its uncertainty using a correlation-function method, finding a lower value, and with a much larger uncertainty, than others have claimed using the same data.  We also explore other methods for estimating the effect of the systematics mentioned and attempt to provide a more realistic error budget to the estimated dipole.

\section{The quasar sample}
\label{sec:WISEquasars}

\subsection{Selecting quasars from WISE}
\label{sec:quasarselection}
\textsc{CatWISE} \citep{catWISE2020} is a comprehensive catalogue of sources selected from WISE \citep{WISE2010} data. WISE was a space mission that conducted an all-sky survey in four infrared bands, detecting a wide range of celestial objects, dominated by stars, but also including millions of extragalactic objects. \textsc{CatWISE} selected sources specifically from the W1~(\SI{3.4}{\micro\meter}) and W2~(\SI{4.6}{\micro\meter}) bands of WISE. 
Secrest et. al.\,\cite{SecrestEtAl2021,SecrestEtAl2022} used a cut on those bands to select a sample of objects with a high probability of being quasars.

\begin{figure*}
    \centering
    \begin{minipage}{0.5\linewidth}
    \includegraphics[width=1\linewidth]{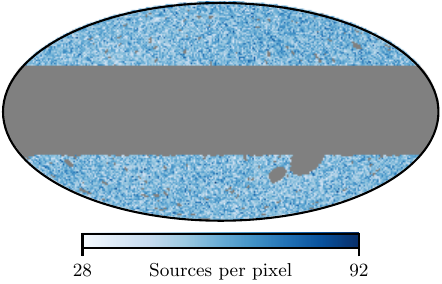}
    \end{minipage}%
    \begin{minipage}{0.5\linewidth}
    \includegraphics[width=1\linewidth]{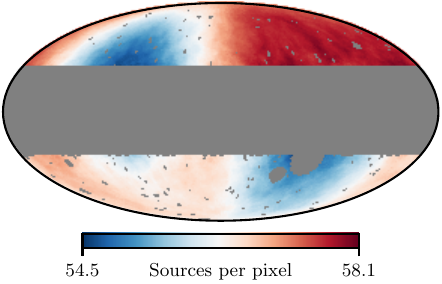}
    \end{minipage}
    \caption{
        \textit{Left:} density map of quasars selected from \textsc{CatWISE}, with the masked region shown in grey.  
        \textit{Right:} same map smoothed with a top-hat filter of area \SI{1}{\steradian} and with the colour scale chosen to show the contrast relative to the average density. The white colour represents the average of the map. 
        }
    \label{fig:quasar_map}
\end{figure*}

We have followed the selection method described by Secrest et al.~\cite{SecrestEtAl2021}, which involves a single \textsc{CatWISE} colour cut, $\mathrm{W1}-\mathrm{W2} \geq 0.8$, to obtain a sample of quasar candidates. For this sample, we found average values of $\alpha = 1.26$ and $x = 1.7$. We made a correction using a Galactic extinction map to select quasars uniformly over the sky. 
Nevertheless, due to the high stellar density near the Galactic plane, a \SI{30}{\degree} cut is applied to mitigate confusion, as described in Secrest et al.\cite{SecrestEtAl2022}. 
Additionally, we exclude certain nearby sources, resulting in the removal of \SI{52.6}{\percent} of the sky. The mask can be seen in grey in \cref{fig:quasar_map}. Once these steps are followed we end up with a catalogue of \num{1355352} probable quasars, matching the number selected by Secrest et al~\cite{SecrestEtAl2021}. The number of quasars per unit of solid angle can then be calculated. To do so we explicitly create a map
using \texttt{HEALPix}\footnote{\url{http://healpix.sourceforge.net}} \citep{gorski2005}, with $N_\mathrm{side}=64$. In \cref{fig:quasar_map} we present areal density plots, with the left-hand panel showing the overall density, while the right-hand panel has been smoothed with a top-hat filter of area \SI{1}{\steradian} and the colour scale has been set to emphasise variations relative to the average density.

Looking at \cref{fig:quasar_map} we note firstly that the quasar distribution on the sky is far from a pure dipole, with more complicated structure being quite apparent. The highest-density direction is \emph{not} opposite the lowest-density direction;  moreover, the map does not show a monotonic gradient over the sky. This realisation leads to an extra correction for the quasar density in the Secrest et al. analysis~\cite{SecrestEtAl2021}, which we describe in more detail in the next section. This map also shows structures on smaller scales; since this structure cannot be due to our motion, then it is presumably due to a combination of Poisson fluctuations (see \cref{sec:simulating_realistic_maps}) in the quasar density and selection effects in the \textsc{CatWISE} data or the creation of the quasar catalogue. We will further investigate this below.

\subsection{Ecliptic gradient}
\label{sec:eclipticgradient}
\begin{figure*}
    \centering
    \begin{minipage}{0.5\linewidth}
    \includegraphics[width=1\linewidth]{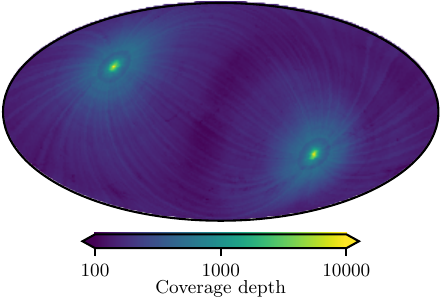}
    \end{minipage}%
    \begin{minipage}{0.5\linewidth}
    \includegraphics[width=1\linewidth]{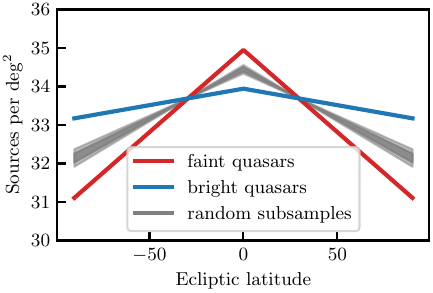}
    \end{minipage}
    \caption{
        \textit{Left:} Coverage across the whole sky in the WISE W1~band; this is a measure of the number of exposures that go into each pixel.  There is considerably more coverage, and hence deeper data, in the regions around the ecliptic poles. 
        \textit{Right:} Quasar density as a function of ecliptic latitude for faint and bright quasars. For the bright half of the quasars, the gradient is almost zero and is about 5 times smaller than for the faint half of the quasars; both of these subsamples give ecliptic gradients that are unusual compared with the gradients for randomly chosen halves of the sample, as indicated by the grey bands, representing the 68\,\% and 95\,\% confidence intervals. This suggests that the quasar data set might be contaminated by stars. 
        }
    \label{fig:CoverageEclipticGradient}
\end{figure*}

\begin{figure}[tbp]
    \centering
    \includegraphics[width=1\linewidth]{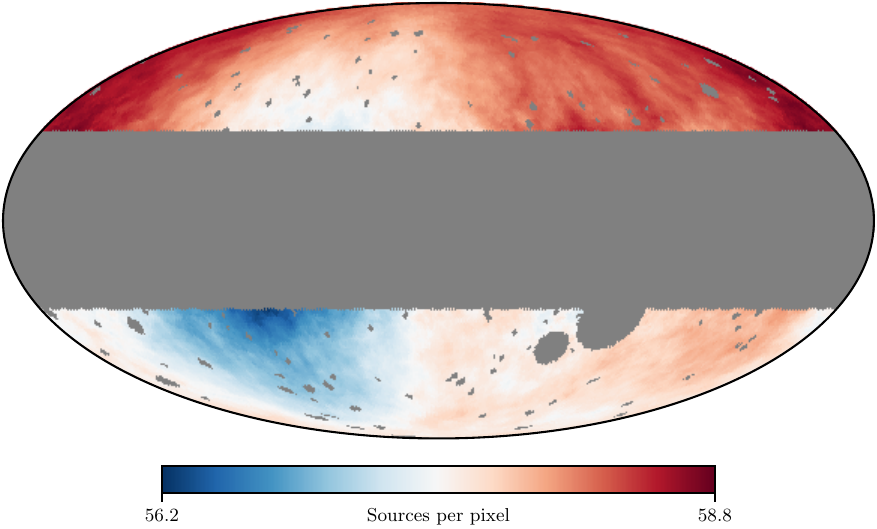}
    \caption{Map of \textsc{CatWISE} quasar density \emph{after} the correction for ecliptic gradient. This map has been smoothed the same way as~\cref{fig:quasar_map}. This map looks quite different from the original map shown in \cref{fig:quasar_map}.  The number of quasars per pixel is higher because in the process of removing the ecliptic gradient, a number of quasars is added to the map. The white colour here represents the average of the map. There certainly appears to be a dipole in this second map, although it is also clear that there is other (higher-multipole) structure.}
    \label{fig:CatWISEMapWOEclipticCorr}
\end{figure}

One important step in the data processing described by Secrest et al.~\cite{SecrestEtAl2021,SecrestEtAl2022}, is the correction for a gradient in ecliptic latitude. This effectively changes the quasar magnitude cut as a function of ecliptic latitude.
We have confirmed that this gradient exists and that it is certainly significant (see the right panel of \cref{fig:CoverageEclipticGradient}, which we will discuss in more detail in \cref{sec:simpletest}). Given the scanning pattern of WISE shown in \cref{fig:CoverageEclipticGradient}, combined with the applied colour cut and potential source confusion, thorough simulations of all of those effects would be necessary to determine if this observed gradient is expected.  For now, the origin of this gradient is unexplained, but
nevertheless, in \cref{fig:CatWISEMapWOEclipticCorr} we show the smoothed quasar-density map after performing this correction.

One other implicit choice made here (seen from \cref{fig:CatWISEMapWOEclipticCorr}) is that the correction for
ecliptic gradient (as carried out by Secrest et al.\,\cite{SecrestEtAl2021}) adds a substantial number of quasars. Before the correction the total is \num{1355352}, and afterwards it is \num{1424517}.  Even if the ecliptic gradient were well understood, it seems unclear that adding quasars is the right way to make this correction, since it changes the monopole of the quasar-density field, complicating the interpretation of the results.Moreover, since the origin of the ecliptic gradient is unexplained, and it can affect the dipole estimate, it is unclear whether removing the gradient is the right thing to do.  We tested the effect of subtracting various ecliptic gradients and found that reasonable choices can change the amplitude of the dipole by around 10\,\% and also shift the direction.  While the ecliptic gradient does not dominate the uncertainty in the final dipole result (but see the estimates in \cref{sec:quasarpowerspectrum}), it is evidence for the presence of systematic effects (presumably caused by the quasar selection process) that are not yet understood, and would have to be fully explained before having confidence in the large-scale properties of the quasar density map. 
Regardless, we apply the same ecliptic gradient correction as in Secrest et al.\
for all subsequent analyses.

\subsection{Stellar contamination}
\label{sec:simpletest}
\begin{figure}
    \centering
    \includegraphics[width= 1\linewidth]{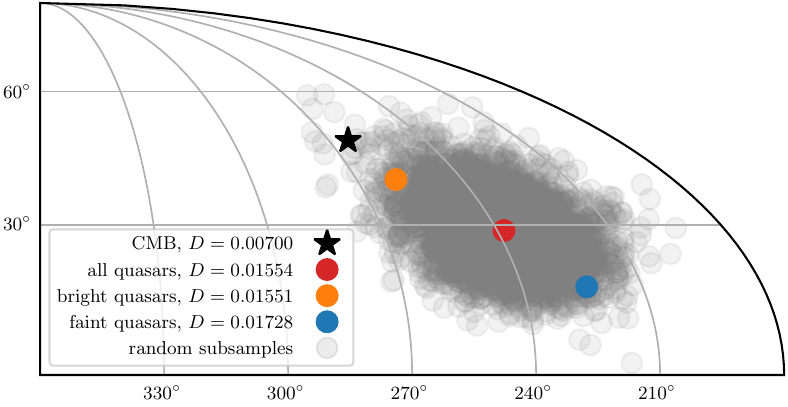}
    \caption{Dipole for the brighter and fainter halves of the quasar sample. Both the direction and the magnitude of the brighter subsample align better with the CMB dipole, compared with the fainter subsample. This suggests that the full sample is contaminated in some way.}
    \label{fig:faint_bright}
\end{figure}

The selection method described in \cref{sec:quasarselection}, and the corrections to the ecliptic gradient discussed in \cref{sec:eclipticgradient}, suggest that the quasar sample may be contaminated by stars. Because there is already a large mask applied to the data to remove the worst of this contamination in the Galactic plane, we should also check that outside of the Galactic plane we have a nearly uncontaminated sample.
To test whether the dipole result obtained in Secrest et al.~\cite{SecrestEtAl2021} could be attributed to selection effects in constructing the quasar sample, we performed a simple test. We divided the quasar sample into two equal subsets based on their brightness in the W1 band (the cut-off magnitude between them is 15.92), and estimated the dipole, using the estimator used in the original study, for these two subsets. In order to determine how significant the bright-versus-faint split results are, we also randomly divided the data into half-size subsamples and analysed those results, which are shown in\cref{fig:CoverageEclipticGradient,fig:faint_bright}.

The results for the dipole direction and amplitude are presented in \cref{fig:faint_bright}. For the brighter subsample, the dipole is $9.5^\circ$ away from the CMB dipole, while for the faint quasars, it is $44.5^\circ$ away. As can be seen, the directions are significantly different from the random cuts in the data (and the size of the shift in direction between the bright and faint halves is seen in less than 1\,\% of random splits). Additionally, the dipole magnitude of the brighter quasars is slightly closer to the CMB dipole.  
These findings suggest that a selection effect, such as stellar contamination within the quasar sample, affects the extracted dipole. We expect that the brighter quasar candidates are more likely to be quasars, with less contamination from the much larger number of faint stars. In a similar vein, we also tested different mask sizes and found that the dipole direction and amplitude changed substantially, depending on how much of the plane of the Milky Way is removed, again suggesting contamination of the quasar sample by stars.

Additionally, as part of the data analysis procedure, we calculated the ecliptic gradient for both sub-samples, as shown in the right panel of \cref{{fig:CoverageEclipticGradient}}. The ecliptic gradient for the bright quasars is significantly smaller than for the faint quasars and is essentially negligible. The ecliptic gradients of the bright and faint subsamples are more extreme compared to randomly-selected subsamples, further indicating that there is some problem with the large-scale properties of the quasar-candidate sample.
In \cref{sec:eclipticgradient} we discussed how the ecliptic gradient of the whole quasar sample has a surprising sign, indicating that it is a systematic effect of some sort.  Checking how the ecliptic gradient changes with quasar brightness further suggests that this is caused by stellar contamination, and not related to the scanning strategy at all.  Until this large-scale selection effect is better understood, we cannot be confident about any analysis of the real large-angle distribution of the quasars.  In future analyses, there may be an opportunity to conduct a comprehensive examination independently on various bright and faint cuts. However, the current sample of quasar candidates is already too limited in number to support such an analysis. 

\section{Masking and high-order moments}
\label{sec:multipoleleakage}
\subsection{How mask-coupling can bias the dipole}
\label{sec:estimators}
We now turn to the second major issue with the existing quasar sample, which comes from the difficulty of estimating the dipole when a large mask is applied to the sky.  For the CMB dipole estimate, the mask covers a relatively small fraction of the sky and the $\ell=1$ mode is much larger than any other multipole.  As we will see, things are much more challenging with the quasars.

The estimator employed by Secrest et al.~\cite{SecrestEtAl2021, SecrestEtAl2022} in their studies is based on a linear regression method. It essentially involves finding the best-fit values for the coefficients of a monopole plus dipole template, by minimizing 
\begin{equation}
\sum_p\left[n_p-\left(\hat{a}_{00}Y_{00}+\hat{a}_{10}Y_{10}+\hat{a}_{11}Y_{11}\right)\right]^2.
\end{equation}
Here $n_p$ is the number of quasars in each pixel, $Y_{\ell m}$ are spherical harmonics and $a_{\ell m}$ are the coefficients in complex number form. Here $\hat{a}$ denotes the fitted values and $a$ represents the true values. With the fitted values for $\hat{a}_{\ell m}$ we can subsequently calculate the power spectrum\footnote{Note that we define $C_\ell$ to be the power spectrum of a particular sky map, not an ensemble-average quantity, as is common elsewhere in the CMB literature.} and the dipole magnitude according to
\begin{equation}
    \label{eq:tildeCls}
    C_\ell\equiv \frac{1}{2\ell+1}\sum_m |\hat{a}_{\ell m}|^2 ;
\end{equation}
and 
\begin{equation}
\label{eq:DC1}
    D = 3 \; \sqrt{\dfrac{C_1}{C_0}}. 
\end{equation}
It is important to remember that $D$ is defined relative to the average density of quasars on the sky (see~\cref{eq:qdipole}), and hence we need to divide $C_1$ by the monopole in the quasar map, $C_0$, in order to interpret the amplitude of the dipole. 

The method we describe above is the same as that used in the \texttt{healpy}\footnote{\url{https://healpy.readthedocs.io}} \texttt{fit\_dipole} function, operating under the assumption that the map is a pure dipole plus independent Gaussian noise. However, as discussed earlier, this assumption is not valid here, since there are clearly other multipole moments in the quasar-density map. 

\begin{figure}
\centering
    \includegraphics[width=1\linewidth]{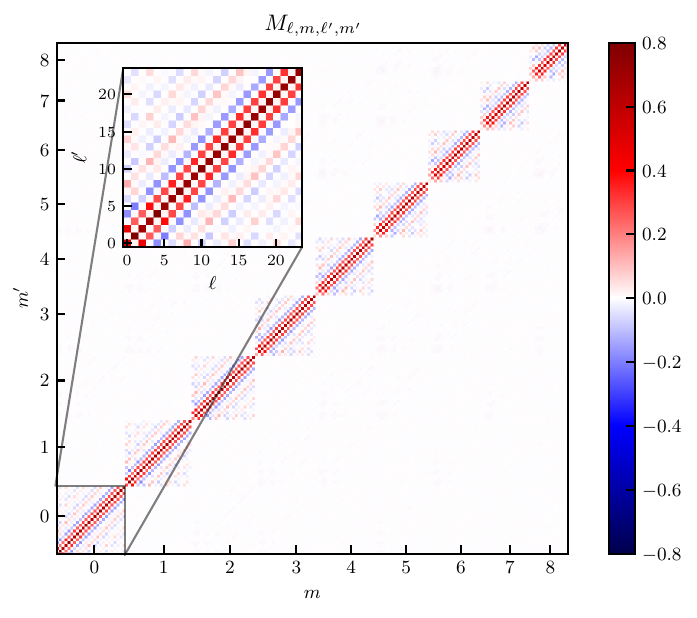}
    \caption{Coupling matrix of spherical harmonic coefficients when the quasar mask is applied. The ordering of the $a_{\ell m}$s is according to the \texttt{HEALPix} package ordering (grouped by $m$ values). We can see that coefficients with the same $m$ values are coupled. The inset plot is a blown-up region, showing the coupling matrix for $m$ and $m'=0$. Notably, the spherical harmonic coefficients $a_{\ell,m}$ exhibit strong coupling with $a_{\ell\pm 2,m}$ due to the resemblance between the Galactic mask and a pure $m=0$ quadrupole.
    }
    \label{fig:coupling_matrix}
\end{figure}

In general, if a regression model fails to account for variables correlated with the existing coefficients, the estimator becomes biased. Due to the substantial masking of the sky, the orthogonality of the spherical harmonic coefficients is compromised, leading to coupling between higher multipoles and dipole coefficients, as detailed in \cref{sec:appendidx}. The coupling matrix $\mathbf{M}$ depends on the size and shape of the mask. The coupling matrix corresponding to the mask used in this study is depicted in \cref{fig:coupling_matrix}, showing a checkerboard pattern. Knowing the coupling between the coefficients one can show that the bias introduced by higher multipoles is given by

\begin{align}
\begin{split}
    \hat{a}_{10} - {a}_{10}  &= \!\!\sum_{\substack{\ell ,m\\ \{\ell, m\} \ne \{1,0\}}} \!\!\!\!\dfrac{M_{\ell,m,1,0}}{M_{1,0,1,0}}a_{\ell m}, \\ 
    \hat{a}_{11} - {a}_{11}  &= \!\!\sum_{\substack{\ell ,m\\ \{\ell, m\} \ne \{1,1\}}}\!\!\!\!\dfrac{M_{\ell,m,1,1}}{M_{1,1,1,1}}a_{\ell m},
\end{split}
\end{align}
where the sum is over all coefficients other than the one being estimated. 

Multiplying the sky by a pattern corresponding to some multipole $\ell_*$ strongly couples modes with $|\Delta\ell|=\ell_*$. Since the Galactic mask has a very strong component of $Y_{2\,0}$, the mask will strongly couple modes with $|\Delta\ell|=2$ (a result familiar from quantum mechanics, and see Ref.~\cite{Moss2011} for the analogous behaviour with $|\Delta\ell|=1$ coupling for dipole modulation), i.e.\ neighbouring even modes and neighbouring odd modes will be strongly connected.  The mask has approximate azimuthal symmetry so spherical harmonics with the
same order~$m$ and even $|\Delta\ell|>2$ are also coupled with each other, as shown in \cref{fig:coupling_matrix}. Therefore, assuming that the dipole is mainly mixed with the odd modes and neglecting other couplings, we can compute the bias introduced by other coefficients (for the particular case of the quasar mask shown in \cref{fig:CatWISEMapWOEclipticCorr}):
\begin{align}
\begin{split}
    \hat{a}_{10}-{a}_{10} &= 
    0.266 \ a_{30} - 0.209 \ a_{50} + 0.067 \ a_{70}+ \dots\ ,
    \\ 
    \hat{a}_{11}-{a}_{11} &= 
    1.296 \ a_{31} - 0.275 \ a_{51} -0.148 \ a_{71} +\dots\ .
\end{split}
\end{align}

One could use a more generalized maximum likelihood method that fits for all spherical harmonic coefficients up to some maximum $\ell$, by minimizing
\begin{equation}
\sum_p\left[n_p-\left(\hat{a}_{00}Y_{00}+\hat{a}_{10}Y_{10}+\hat{a}_{11}Y_{11}+\hat{a}_{20}Y_{20}+\hat{a}_{21}Y_{21}+\dots+\hat{a}_{\ell_{\mathrm{max}}m_{\mathrm{max}}}Y_{\ell_{\mathrm{max}}m_{\mathrm{max}}}\right)\right]^2.
\end{equation}
However, in the presence of a large mask, this model becomes very unstable, leading to inflated coefficient estimates. This is similar to the situation of trying to estimate the quadrupole in the early days of analysis of Cosmic Background Explorer (\textit{COBE}) satellite data.  For instance, Bunn et al.\,\cite{Bunn1994} concluded that ``it is in general impossible to extract reliable information about particular multipoles from an incomplete data set; for example, attempts to determine the local quadrupole moment of the CMB anisotropy from the COBE data set will inevitably have large uncertainties''.  The current situation with the dipole is more difficult still, given the very large mask that is required.

\begin{figure*}
\includegraphics[width = 1\linewidth]{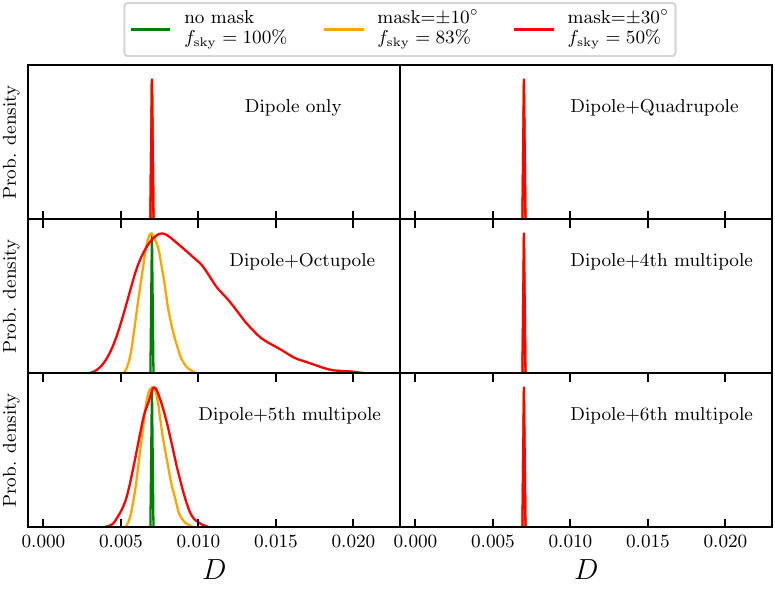}
  \caption{Quasar maps simulated with the null hypothesis that their dipole is equal to the CMB dipole, $D^\mathrm{input}=0.007$. In each of the $\ell$th panels, multipole $\ell$ is added, with the same power as the dipole (i.e.\ $C_\ell = C_1^\mathrm{input}$, see \cref{eq:tildeCls}) and without cosmic variance. The histograms of estimated dipoles for these maps are shown in different colours according to the applied mask. The red histograms are computed after applying a mask similar to the one used in this study, while the green histograms show the computed dipole without applying any mask, and in orange, we show the results for a mask of intermediate size. As can be seen, the addition of higher multipoles, especially the octupole, can have large effects on the dipole estimation when the mask is large.
  }
\label{fig:SimulatedQuasarMapDipoleHistogram}
\end{figure*}

To see this in action and compare the current mask with smaller masks, we present a series of simulations in \cref{fig:SimulatedQuasarMapDipoleHistogram}.  For the sake of illustration, we choose multipoles such that the dipole has the CMB value and then combine with other multipoles one by one, each having observed $C_\ell=C_1$ with randomly chosen orientations. Here $C_\ell$ is the power spectrum estimated from a single map (without cosmic variance).
We generate these simulations using the \texttt{synalm} function of \texttt{healpy}, and we scale the $a_{\ell m}$ coefficients such that they produce the desired~$C_\ell$ exactly (i.e.\ without cosmic variance);
after making each of these simulations we apply the quasar mask and determine the dipole with \texttt{healpy}'s \texttt{fit\_dipole}.  The first panel shows $\ell=1$ only, the second panel adds random quadrupoles with $C_2=C_1$, the third panel adds random octupoles with $C_3=C_1$, etc. The green distribution illustrates the dipole estimate when there is no mask applied, which recovers the input dipole ($D=0.007$). 
We notice from the figure that the presence of a non-negligible octupole can substantially alter the dipole estimate of the \texttt{fit\_dipole} function; this is because of the strong $|\Delta\ell|=2$ coupling caused by the mask.

\subsection{Simulating mask-induced effects}
\label{sec:simulating_realistic_maps}

In Secrest et al.~\cite{SecrestEtAl2021}, the $p$-value for measuring a dipole of the quasar map that is twice as large as the CMB dipole within the framework of $\Lambda$CDM is computed as follows. A series of quasar maps with the CMB dipole as input is simulated and the $p$-value is defined as the fraction of simulated skies with dipole amplitudes exceeding the quasar dipole on the real sky. Here, we perform a similar analysis, only this time we want to include higher multipoles that are similar to those that might be in the quasar data set.

\begin{figure*}
    \includegraphics[width=\textwidth]{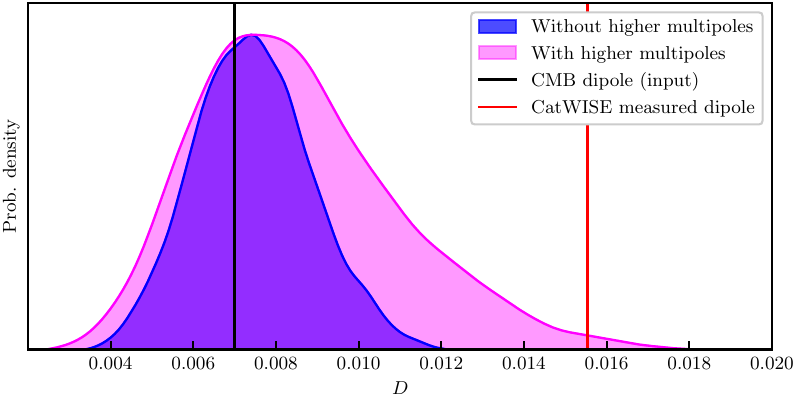}
    \caption{Dipole estimates of quasar mock maps generated with the CMB dipole (black line) as input, and with randomly oriented quadrupoles and octupoles. These higher multipoles are all constrained to be smaller than the dipole. We estimate the dipole magnitude here using \texttt{fit\_dipole} after masking the maps. These normalized probability densities show how the inclusion of higher-order multipoles (the odd multipoles, to be more specific, see \cref{fig:coupling_matrix,fig:SimulatedQuasarMapDipoleHistogram}) biases the dipole estimator. The uncertainty in the blue histogram is due to Poisson noise and masking, which is not enough to explain the measured quasar dipole. However, the presence of systematic higher multipoles would increase the probability of such measurement substantially.}
    \label{fig:QuasarSimWithSyst}
\end{figure*}

There is an additional effect for this type of simulation due to the Poisson nature of the quasar number counts, which also contributes to the uncertainty in the dipole. In order to take that into account, this time we do not use \texttt{synalm}. Instead, we simulate quasar maps in pixel space with the same number of quasars using a Poisson distribution. Subsequently, we modulate these maps by the CMB dipole. 
We generate two sets of these maps, each comprising 10\,000 maps. One set is intended to mimic ideal quasar maps unaffected by selection effects. The blue histogram in~\cref{fig:QuasarSimWithSyst} shows the distribution of the dipole magnitude estimated from these maps. 
This is similar to figure~4 in Secrest et al.~\cite{SecrestEtAl2021} Based on these mock maps, the null hypothesis seems to be ruled out by almost $5\,\sigma$.

For the other set of mock maps, we introduce higher multipoles to simulate selection effects in the quasar sample or other systematic effects on large scales. We adopt a simplifying assumption that the amplitudes of these higher multipoles are all smaller than that of the dipole, and we only consider moments up to $\ell=3$ for simplicity. To generate these higher multipoles in our quasar maps, we create random maps with quadrupoles and octupoles, where their magnitudes are drawn from a uniform distribution smaller than the dipole magnitude. We use the values on the pixels of these maps as weights for our quasar maps, modulating our quasar maps accordingly.
The results, presented in \cref{fig:QuasarSimWithSyst}, show the distribution of the measured dipole magnitude of these quasar maps. In this illustrative example, the $p$-value is $0.014$, corresponding to a $2.2\,\sigma$ significance level for a one-tailed Gaussian distribution. This value is substantially higher than the initial $p$-value calculated under the assumption of the absence of quadrupole or octupole contributions. Keep in mind that this was done only for quadrupole and octupole modes assumed to be of smaller magnitude than the dipole; if we considered higher and stronger modes the $p$-value could decrease further. 

This analysis underscores the importance of considering systematic effects in the spatial distribution of the quasars. These simulations indicate that if the amount of masked sky is forced to be large, then producing a robust dipole estimate requires a quasar sample with negligible higher-order multipoles and much better-understood selection effects.  Additionally, a considerably larger quasar sample would enable a more detailed study of how the dipole in subsamples varies with mask size, quasar brightness, etc., which would hopefully lead to a better understanding and quantification of the systematic effects.

\section{Estimating the quasar map power spectrum}
\label{sec:sim_real_qm}
\subsection{Power spectrum estimators}
\label{sec:powerspectrumestimators}
Since we have shown that higher-order structure in the quasar-density map could affect the estimate of the dipole, one would naturally like to know how big the higher multipoles are in the quasar data.  However, obtaining a robust estimate is challenging, for the same reason as the mask-induced coupling---the 50\,\% mask makes it hard to determine the power spectrum.

There are several different approaches that could be taken to estimate the power spectrum in a masked map.
In \textit{Planck} CMB analyses, the masked region can be filled in using constrained Gaussian realizations, as part of a Gibbs-sampling approach~\cite{Hoffman1991,Eriksen2004}. However, during this process, the monopole and dipole coefficients are usually estimated using template fitting, as described above, while there is an implicit assumption that the other modes have Gaussian statistics. Since the CMB monopole and dipole magnitude are significantly larger than other multipole moments, then severe biases can be avoided by employing template fitting to estimate these modes.  However, for the quasar-density map, there is no reason to believe a priori that the $\ell=0$ and $\ell=1$ modes are orders of magnitude larger than the others (compare \cref{fig:CMBDipole} to \cref{fig:quasar_map}).  Moreover, we cannot assume that the statistics are Gaussian for the $\ell\geq2$ modes on the quasar sky (as they are for CMB maps), and hence it is not clear that one can use the same Gibbs-sampling approach that has been used for CMB power spectrum estimation.

There are many other methods that are used in CMB analyses. We have tested several of them for application to simulated quasar samples; however, none of the methods investigated gives the desired accuracy for estimating the dipole when a 50\,\% mask is applied. Quadratic maximum likelihood (QML) methods are the most commonly used in CMB studies~\cite{Tegmark1997,Tegmark2001}. This approach is optimized to minimize the loss of information by imposing external constraints on the CMB power spectrum. However, when we do not have a reliable prior model for the fiducial power spectrum, this method does not produce reliable results.

Another approach would be to perform a simultaneous estimate of a set of $a_{\ell m}$ modes, including taking account of their correlations (as we did in \cref{sec:estimators}).  When we tried this for the quasar map, we found that the results depend sensitively upon the maximum multipole~$\ell_\mathrm{max}$ that is considered.
Pseudo-$C_\ell$ methods are another family of techniques widely used in CMB studies~\cite{Wandelt2001,Hivon2002}. They offer an approximation to the $C_\ell$s by employing cross-correlations of spherical harmonic coefficients instead of straightforward direct summation. However, a drawback with this method is that it does not directly estimate the dipole and furthermore it is usually applied to binned power-spectrum estimates and, like other methods, does not work well for individual multipoles when there is a large mask. It seems that perhaps the best strategy for large masks is to start from the position-space correlation function, an approach that we describe next.

\subsection{Applying a power spectrum estimator}
\label{sec:quasarpowerspectrum}

As discussed earlier in \cref{sec:quasarselection}, visual inspection of the quasar map (\cref{fig:quasar_map,fig:CatWISEMapWOEclipticCorr}) indicates that it is contaminated with some systematic effects on large scales. However, as explained earlier, measuring the large-scale power in the map is challenging and most conventional methods do not work well.

Though estimating the low-order modes in the quasar density map is challenging, we make an attempt nevertheless, to obtain some kind of estimate of the large-scale modes, in order to place a more realistic uncertainty on the dipole estimate.
The position-space correlation function is unbiased by the application of a mask and provides a well-behaved route to estimating the power spectrum on large scales~\cite{Szapudi2001}.  Therefore, in an attempt to quantify the amplitude of the low-$\ell$ modes in the quasar map, we turn to \texttt{PolSpice}\footnote{\url{https://www2.iap.fr/users/hivon/software/PolSpice/}}~\cite{Szapudi2001,Chon2004}, which uses the Legendre transform of the correlation function $C(\theta)$ to estimate the angular power spectrum. 

\begin{figure*}
    \includegraphics[height=5.15cm, width = 0.42\textwidth]{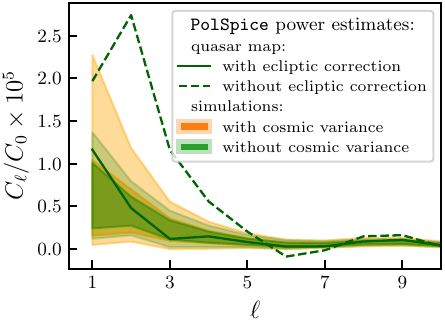}
    \hfill
    \includegraphics[height=5.15cm,width=0.56\textwidth]{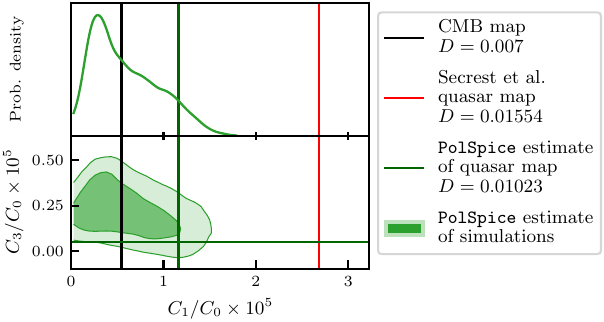}
    \caption{
        \textit{Left:} Power spectrum estimate of the quasar map using \texttt{PolSpice}.  The bands show the 68\,\% and 95\,\% ranges from \num{10000} simulations, both without (green) and with (yellow) cosmic variance.
        The solid line is the best \texttt{PolSpice} estimate, while the dashed line is the same thing if the ecliptic gradient is \emph{not} removed.
        \textit{Right:}~The green lines and contours (plotted using \texttt{anesthetic}~\cite{anesthetic}) show the joint distribution of $C_1$ and $C_3$ and the marginalised probability density of the dipole power~($C_1$)  estimates from simulations, with the quasar map estimates (green vertical line) as input. The simulations were normalised to exactly match the input power spectrum (removing any contribution from cosmic variance). The uncertainty shown here is therefore dominated by masking effects.
        }
\label{fig:PolSpice}
\end{figure*}

The result is shown in \cref{fig:PolSpice}. The measured dipole is about $D=0.010$, with large statistical and systematic uncertainties. The green bands plotted in \cref{fig:PolSpice} show the \SI{68}{\percent} and \SI{95}{\percent} ranges for a set of realisations of the power spectrum given by the solid green line, treated the same way as the data.  Explicitly, we choose random directions but do not pick random amplitudes from a Gaussian (i.e., there is no cosmic variance included here). For each of these realisations, we apply the quasar mask and run \texttt{PolSpice}. The green band gives an indication of the statistical uncertainties inherent in the quasar data, including the mask-induced coupling effects.
The yellow bands show the \SI{68}{\percent} and \SI{95}{\percent} ranges when cosmic variance \emph{is} included. The right panel in \cref{fig:PolSpice} shows the joint probability density of $C_1$ and $C_3$, along with estimated values for the quasar map and the CMB dipole value. It also shows the correlation between the dipole and octupole modes. 

In an ideal quasar density map, we would have a strong dipole plus some small additional higher multipoles caused by Poisson statistics, which would produce $C_\ell=$ constant for all $\ell>1$. However, as can be seen in~\cref{fig:PolSpice}, the magnitude of the power on large scales ($2\leq\ell<10$) is significantly higher than that expected just from Poisson statistics given $C_1$.  The dashed line is the power spectrum \emph{before} removing the ecliptic-latitude gradient, while the solid line shows the estimate \emph{after} the removal.  Since this gradient is unexplained, we cannot say which of the two green lines is more correct, and the difference between them provides some estimate of the size of the systematic uncertainties.  Considering the difference between the solid and dashed lines, and the coloured bands in \cref{fig:PolSpice}, the relative uncertainty in the dipole is of order unity.

In summary, estimating this higher-order structure in the quasar density map is challenging and depends sensitively on what exactly is done.  The results with and without the ecliptic-gradient removal are quite different, as shown in \cref{fig:PolSpice}.  The results depend to some extent on the maximum multipole considered, and additionally, we obtain very different results depending on whether or not we include the dipole in the set of multipoles to be estimated.  This is because of the mask-induced coupling, particularly between the dipole and octupole, as shown in the right panel of \cref{fig:PolSpice}.  The outcome of these efforts to estimate the power spectrum in the quasar map is that the magnitude of the dipole is highly uncertain.

\section{Conclusions}
\label{sec:conclusions}
Studying the large-scale statistical properties of the Universe is certainly a worthwhile endeavour, and many studies have followed the claims made by Secrest et al., suggesting that there might be something genuinely missing in our understanding of the source-generated dipole (e.g.~\cite{Dam2022,Guandalin2023,Singal2024, Mittal2023}).  We believe that most of these studies have missed the most important issue, which is the control of systematics, particularly when one is forced to mask such a large fraction of the sky.

In our study, we aimed to replicate the results from Secrest et al. As we went through this process, we came across several pieces of evidence pointing to the presence of noticeable systematic effects in the data---the data seem to be affected by unexplained selection effects on large scales. These effects cause significant power at very low $\ell$, so until they are thoroughly understood we cannot rule out the possibility that they also have power at $\ell = 1$, making it impossible to do a straightforward comparison of the quasar dipole with the CMB dipole.

We performed a bright-faint split test, where we found that the dipole of the brighter quasars is closer to the CMB dipole. Due to the limitation of the size of the data set, we could not carry out a more detailed segmentation analysis of subsets of the data.  A much larger sample of distant extragalactic sources would be needed to adequately perform such tests.

The second major factor affecting the quasar-density dipole estimate is the necessity of applying a very large mask, with more than \SI{50}{\percent} of the sky removed.  This leads to strong coupling between multipoles and makes it challenging to robustly estimate the dipole.  This situation is very different from the case of the CMB, where the dipole is two orders of magnitude larger than the other multipoles, and the amount of sky removed when estimating the dipole can be as small as \SI{20}{\percent}.

We further investigated how the systematic effects might combine with the large mask to bias the dipole estimates and substantially increase the uncertainties.  We showed that the mask used introduces a non-negligible bias to the dipole, which could only be reduced by ensuring that the octupole was negligible (something we are currently unable to do).  Assuming that the higher-order multipoles in the quasar distribution might be of similar amplitude to the dipole greatly increases the measured uncertainty. Thus, the observed dipole becomes much less discrepant with the expectation from the CMB dipole. 

Lastly, we employed a pixel-space correlation function method to estimate the dipole and higher multipoles of the quasar map at the same time. The resulting dipole determination is smaller than we found when estimating the dipole on its own and has much larger uncertainty.  Although it is hard to be confident in this result (because of the fundamental problem of the large mask causing mode-coupling), this at least suggests that the level of uncertainty coming from the simple dipole estimate has been underestimated.  The uncertainty could be reduced if we could construct a cleaner quasar map, dominated on large scales by only the dipole.

Since the quasar sample investigated in this paper was selected using only a single colour criterion, one obvious improvement would be to include additional data to reduce stellar contamination.  This has already been done through construction of the Quaia catalogue~\cite{Quaia2023}, where Gaia satellite data were combined with WISE to make a cleaner quasar sample.  We looked into this new sample but found that it is not obviously cleaner (at least for the purposes of estimating the dipole); for example, it contains a very strong Galactic-latitude gradient, which will have to be carefully dealt with before estimating the source dipole.  A similar conclusion was also reached in a recent independent study~\cite{Mittal2023}. 

Moving forward, future surveys will need to address these systematic effects to ensure the accuracy and reliability of their results. One key aspect will be the utilization of larger and more comprehensive data sets to conduct thorough analyses. With a larger data set, we will be better equipped to identify and mitigate systematic effects, as well as perform more thorough tests of subsamples to understand and control such effects.

Things would be much simpler if one could simply make a catalogue of extragalactic sources over the entire sky, or at least over the vast majority of the sky.  However, in the optical, the extinction caused by Milky Way dust, plus the dramatic increase in stellar surface density towards the Galactic plane, makes it unrealistic to expect an optical catalogue with a mask that is much smaller than the \SI{50}{\percent} mask studied here.  Perhaps future radio surveys offer more promising prospects for wider coverage of the sky.  However, in all wavebands it will continue to be challenging to confidently measure a difference between one side of the sky and the other of order $10^{-3}$ that is free of systematic effects at a similar level, which is what is required to determine whether the dipole in source counts differs significantly from the CMB dipole.

\acknowledgments
We acknowledge financial support from the Natural Sciences and Engineering Research Council of Canada.  We thank Secrest et al.\ for making their data and code available so that their results could be readily reproduced. We also thank Jim Zibin for helpful comments. LTH was supported through a Killam Postdoctoral Fellowship and a CITA National Fellowship. Some results in this paper were derived using the \texttt{HEALPix} (\url{http://healpix.sourceforge.net}), \texttt{healpy} (\url{https://healpy.readthedocs.io}), \ and \texttt{anesthetic} (\url{https://anesthetic.readthedocs.io}) packages.

\bibliographystyle{JHEP}
\bibliography{Biblio,Planck_bib}

\providecommand{\href}[2]{#2}\begingroup\raggedright\begin{thebibliography}{10}

\bibitem{Peebles1993}
P.J.E.~{Peebles}, \emph{{Principles of Physical Cosmology}}, {Princeton University Press} (1993), \href{https://doi.org/10.1515/9780691206721}{10.1515/9780691206721}.

\bibitem{Peebles1968}
P.J.~{Peebles} and D.T.~{Wilkinson}, \emph{{Comment on the Anisotropy of the Primeval Fireball}}, \href{https://doi.org/10.1103/PhysRev.174.2168}{\emph{Physical Review} {\bfseries 174} (1968) 2168}.

\bibitem{planck2016-l01}
{\sorthelp{Planck Collaboration 2018A}}{Planck Collaboration I}, \emph{{\textit{Planck} 2018 results. I. Overview, and the cosmological legacy of \textit{Planck}}}, \href{https://doi.org/10.1051/0004-6361/201833880}{\emph{\aap} {\bfseries 641} (2020) A1} [\href{https://arxiv.org/abs/1807.06205}{{\ttfamily 1807.06205}}].

\bibitem{planck2016-l03}
{}{Planck Collaboration III}, \emph{{\textit{Planck} 2018 results. III. High Frequency Instrument data processing}}, \href{https://doi.org/10.1051/0004-6361/201832909}{\emph{\aap} {\bfseries 641} (2020) A3} [\href{https://arxiv.org/abs/1807.06207}{{\ttfamily 1807.06207}}].

\bibitem{planck2013-pipaberration}
{\sorthelp{Planck Collaboration 2014ZB}}{Planck Collaboration XXVII}, \emph{{\textit{Planck} 2013 results. XXVII. Doppler boosting of the CMB: Eppur si muove}}, \href{https://doi.org/10.1051/0004-6361/201321556}{\emph{\aap} {\bfseries 571} (2014) A27} [\href{https://arxiv.org/abs/1303.5087}{{\ttfamily 1303.5087}}].

\bibitem{planck2020-LVI}
{\sorthelp{Planck Collaboration IntZZF}}{Planck Collaboration Int. LVI}, \emph{{\textit{Planck} intermediate results. LVI. Detection of the CMB dipole through modulation of the thermal Sunyaev-Zeldovich effect: Eppur si muove II}}, \href{https://doi.org/10.1051/0004-6361/202038053}{\emph{\aap} {\bfseries 644} (2020) 100} [\href{https://arxiv.org/abs/2003.12646}{{\ttfamily 2003.12646}}].

\bibitem{Saha2021}
S.~{Saha}, S.~{Shaikh}, S.~{Mukherjee}, T.~{Souradeep} and B.D.~{Wandelt}, \emph{{Bayesian estimation of our local motion from the Planck-2018 CMB temperature map}}, \href{https://doi.org/10.1088/1475-7516/2021/10/072}{\emph{\jcap} {\bfseries 2021} (2021) 072} [\href{https://arxiv.org/abs/2106.07666}{{\ttfamily 2106.07666}}].

\bibitem{MRR2000}
M.~{Rowan-Robinson}, J.~{Sharpe}, S.J.~{Oliver}, O.~{Keeble}, A.~{Canavezes}, W.~{Saunders} et~al., \emph{{The IRAS PSCz dipole}}, \href{https://doi.org/10.1046/j.1365-8711.2000.03313.x}{\emph{\mnras} {\bfseries 314} (2000) 375} [\href{https://arxiv.org/abs/astro-ph/9912223}{{\ttfamily astro-ph/9912223}}].

\bibitem{planck2013-XIII}
{\sorthelp{Planck Collaboration IntM}}{Planck Collaboration Int. XIII}, \emph{{\textit{Planck} intermediate results. XIII. Constraints on peculiar velocities}}, \href{https://doi.org/10.1051/0004-6361/201321299}{\emph{\aap} {\bfseries 561} (2014) A97} [\href{https://arxiv.org/abs/1303.5090}{{\ttfamily 1303.5090}}].

\bibitem{Erdogdu2006}
P.~{Erdo{\v{g}}du}, J.P.~{Huchra}, O.~{Lahav}, M.~{Colless}, R.M.~{Cutri}, E.~{Falco} et~al., \emph{{The dipole anisotropy of the 2 Micron All-Sky Redshift Survey}}, \href{https://doi.org/10.1111/j.1365-2966.2006.10243.x}{\emph{\mnras} {\bfseries 368} (2006) 1515} [\href{https://arxiv.org/abs/astro-ph/0507166}{{\ttfamily astro-ph/0507166}}].

\bibitem{Turner1991}
M.S.~Turner, \emph{{A Tilted Universe (and Other Remnants of the Preinflationary Universe)}}, \href{https://doi.org/10.1103/PhysRevD.44.3737}{\emph{Phys. Rev. D} {\bfseries 44} (1991) 3737}.

\bibitem{Turner1992}
M.S.~{Turner}, \emph{{The tilted universe}}, \href{https://doi.org/10.1007/BF00756869}{\emph{General Relativity and Gravitation} {\bfseries 24} (1992) 1}.

\bibitem{ZibinScott2008}
J.P.~{Zibin} and D.~{Scott}, \emph{{Gauging the cosmic microwave background}}, \href{https://doi.org/10.1103/PhysRevD.78.123529}{\emph{\prd} {\bfseries 78} (2008) 123529} [\href{https://arxiv.org/abs/0808.2047}{{\ttfamily 0808.2047}}].

\bibitem{EllisBaldwin1984}
G.F.R.~{Ellis} and J.E.~{Baldwin}, \emph{{On the expected anisotropy of radio source counts}}, \href{https://doi.org/10.1093/mnras/206.2.377}{\emph{\mnras} {\bfseries 206} (1984) 377}.

\bibitem{MeiksinDavis1986}
A.~{Meiksin} and M.~{Davis}, \emph{{Anisotropy of the galaxies detected by IRAS}}, \href{https://doi.org/10.1086/113999}{\emph{\aj} {\bfseries 91} (1986) 191}.

\bibitem{Lahav1987}
O.~{Lahav}, \emph{{Optical dipole anisotropy}}, \href{https://doi.org/10.1093/mnras/225.2.213}{\emph{\mnras} {\bfseries 225} (1987) 213}.

\bibitem{Scharf2000}
C.A.~{Scharf}, K.~{Jahoda}, M.~{Treyer}, O.~{Lahav}, E.~{Boldt} and T.~{Piran}, \emph{{The 2-10 keV X-Ray Background Dipole and Its Cosmological Implications}}, \href{https://doi.org/10.1086/317174}{\emph{\apj} {\bfseries 544} (2000) 49} [\href{https://arxiv.org/abs/astro-ph/9908187}{{\ttfamily astro-ph/9908187}}].

\bibitem{Watkins2023}
R.~Watkins, T.~Allen, C.J.~Bradford, A.~Ramon, A.~Walker, H.A.~Feldman et~al., \emph{Analysing the large-scale bulk flow using cosmicflows4: increasing tension with the standard cosmological model}, \href{https://doi.org/10.1093/mnras/stad1984}{\emph{Monthly Notices of the Royal Astronomical Society} {\bfseries 524} (2023) 1885–1892}.

\bibitem{Tiwari2024}
P.~Tiwari, D.J.~Schwarz, G.-B.~Zhao, R.~Durrer, M.~Kunz and H.~Padmanabhan, \emph{{An Independent Measure of the Kinematic Dipole from SDSS}}, {\emph{arXiv} (2024) } [\href{https://arxiv.org/abs/2409.09946}{{\ttfamily 2409.09946}}].

\bibitem{BlakeWall2002}
C.~{Blake} and J.~{Wall}, \emph{{A velocity dipole in the distribution of radio galaxies}}, \href{https://doi.org/10.1038/416150a}{\emph{\nat} {\bfseries 416} (2002) 150} [\href{https://arxiv.org/abs/astro-ph/0203385}{{\ttfamily astro-ph/0203385}}].

\bibitem{Crawford2009}
F.~{Crawford}, \emph{{Detecting the Cosmic Dipole Anisotropy in Large-Scale Radio Surveys}}, \href{https://doi.org/10.1088/0004-637X/692/1/887}{\emph{\apj} {\bfseries 692} (2009) 887} [\href{https://arxiv.org/abs/0810.4520}{{\ttfamily 0810.4520}}].

\bibitem{Gibelyou2012}
C.~{Gibelyou} and D.~{Huterer}, \emph{{Dipoles in the sky}}, \href{https://doi.org/10.1111/j.1365-2966.2012.22032.x}{\emph{\mnras} {\bfseries 427} (2012) 1994} [\href{https://arxiv.org/abs/1205.6476}{{\ttfamily 1205.6476}}].

\bibitem{Rubart2014}
M.~{Rubart}, D.~{Bacon} and D.J.~{Schwarz}, \emph{{Impact of local structure on the cosmic radio dipole}}, \href{https://doi.org/10.1051/0004-6361/201423583}{\emph{\aap} {\bfseries 565} (2014) A111} [\href{https://arxiv.org/abs/1402.0376}{{\ttfamily 1402.0376}}].

\bibitem{TiwariNusser2016}
P.~{Tiwari} and A.~{Nusser}, \emph{{Revisiting the NVSS number count dipole}}, \href{https://doi.org/10.1088/1475-7516/2016/03/062}{\emph{\jcap} {\bfseries 2016} (2016) 062} [\href{https://arxiv.org/abs/1509.02532}{{\ttfamily 1509.02532}}].

\bibitem{Darling_2022}
J.~Darling, \emph{The universe is brighter in the direction of our motion: Galaxy counts and fluxes are consistent with the {CMB} dipole}, \href{https://doi.org/10.3847/2041-8213/ac6f08}{\emph{The Astrophysical Journal Letters} {\bfseries 931} (2022) L14}.

\bibitem{Wagenveld2023}
J.D.~{Wagenveld}, H.R.~{Kl{\"o}ckner} and D.J.~{Schwarz}, \emph{{The cosmic radio dipole: Bayesian estimators on new and old radio surveys}}, \href{https://doi.org/10.1051/0004-6361/202346210}{\emph{\aap} {\bfseries 675} (2023) A72} [\href{https://arxiv.org/abs/2305.15335}{{\ttfamily 2305.15335}}].

\bibitem{Wagenveld2024}
J.D.~Wagenveld et~al., \emph{{The MeerKAT Absorption Line Survey Data Release 2: Wideband continuum catalogues and a measurement of the cosmic radio dipole}}, {\emph{arXiv} (2024) } [\href{https://arxiv.org/abs/2408.16619}{{\ttfamily 2408.16619}}].

\bibitem{Singal_2011}
A.K.~{Singal}, \emph{{Large Peculiar Motion of the Solar System from the Dipole Anisotropy in Sky Brightness due to Distant Radio Sources}}, \href{https://doi.org/10.1088/2041-8205/742/2/L23}{\emph{\apjl} {\bfseries 742} (2011) L23} [\href{https://arxiv.org/abs/1110.6260}{{\ttfamily 1110.6260}}].

\bibitem{Rubart_2013}
M.~{Rubart} and D.J.~{Schwarz}, \emph{{Cosmic radio dipole from NVSS and WENSS}}, \href{https://doi.org/10.1051/0004-6361/201321215}{\emph{\aap} {\bfseries 555} (2013) A117} [\href{https://arxiv.org/abs/1301.5559}{{\ttfamily 1301.5559}}].

\bibitem{Colin2017}
J.~{Colin}, R.~{Mohayaee}, M.~{Rameez} and S.~{Sarkar}, \emph{{High-redshift radio galaxies and divergence from the CMB dipole}}, \href{https://doi.org/10.1093/mnras/stx1631}{\emph{\mnras} {\bfseries 471} (2017) 1045} [\href{https://arxiv.org/abs/1703.09376}{{\ttfamily 1703.09376}}].

\bibitem{Tiwari_2019}
P.~Tiwari and P.K.~Aluri, \emph{Large angular-scale multipoles at redshift $\sim$ 0.8}, \href{https://doi.org/10.3847/1538-4357/ab1d58}{\emph{The Astrophysical Journal} {\bfseries 878} (2019) 32}.

\bibitem{Singal2019}
A.K.~{Singal}, \emph{{Large disparity in cosmic reference frames determined from the sky distributions of radio sources and the microwave background radiation}}, \href{https://doi.org/10.1103/PhysRevD.100.063501}{\emph{\prd} {\bfseries 100} (2019) 063501} [\href{https://arxiv.org/abs/1904.11362}{{\ttfamily 1904.11362}}].

\bibitem{Siewert2020}
T.M.~Siewert, M.~Schmidt-Rubart and D.J.~Schwarz, \emph{{Cosmic radio dipole: Estimators and frequency dependence}}, \href{https://doi.org/10.1051/0004-6361/202039840}{\emph{Astron. Astrophys.} {\bfseries 653} (2021) A9} [\href{https://arxiv.org/abs/2010.08366}{{\ttfamily 2010.08366}}].

\bibitem{Oayda2024}
O.T.~{Oayda}, V.~{Mittal}, G.F.~{Lewis} and T.~{Murphy}, \emph{{A Bayesian approach to the cosmic dipole in radio galaxy surveys: joint analysis of NVSS \& RACS}}, \href{https://doi.org/10.1093/mnras/stae1399}{\emph{\mnras} {\bfseries 531} (2024) 4545} [\href{https://arxiv.org/abs/2406.01871}{{\ttfamily 2406.01871}}].

\bibitem{SecrestEtAl2021}
N.J.~{Secrest}, S.~{von Hausegger}, M.~{Rameez}, R.~{Mohayaee}, S.~{Sarkar} and J.~{Colin}, \emph{{A Test of the Cosmological Principle with Quasars}}, \href{https://doi.org/10.3847/2041-8213/abdd40}{\emph{\apjl} {\bfseries 908} (2021) L51} [\href{https://arxiv.org/abs/2009.14826}{{\ttfamily 2009.14826}}].

\bibitem{SecrestEtAl2022}
N.J.~{Secrest}, S.~{von Hausegger}, M.~{Rameez}, R.~{Mohayaee} and S.~{Sarkar}, \emph{{A Challenge to the Standard Cosmological Model}}, \href{https://doi.org/10.3847/2041-8213/ac88c0}{\emph{\apjl} {\bfseries 937} (2022) L31} [\href{https://arxiv.org/abs/2206.05624}{{\ttfamily 2206.05624}}].

\bibitem{catWISE2020}
P.R.M.~{Eisenhardt}, F.~{Marocco}, J.W.~{Fowler}, A.M.~{Meisner}, J.D.~{Kirkpatrick}, N.~{Garcia} et~al., \emph{{The CatWISE Preliminary Catalog: Motions from WISE and NEOWISE Data}}, \href{https://doi.org/10.3847/1538-4365/ab7f2a}{\emph{\apjs} {\bfseries 247} (2020) 69} [\href{https://arxiv.org/abs/1908.08902}{{\ttfamily 1908.08902}}].

\bibitem{WISE2010}
E.L.~{Wright}, P.R.M.~{Eisenhardt}, A.K.~{Mainzer}, M.E.~{Ressler}, R.M.~{Cutri}, T.~{Jarrett} et~al., \emph{{The Wide-field Infrared Survey Explorer (WISE): Mission Description and Initial On-orbit Performance}}, \href{https://doi.org/10.1088/0004-6256/140/6/1868}{\emph{\aj} {\bfseries 140} (2010) 1868} [\href{https://arxiv.org/abs/1008.0031}{{\ttfamily 1008.0031}}].

\bibitem{Das2021}
K.K.~{Das}, K.~{Sankharva} and P.~{Jain}, \emph{{Explaining excess dipole in NVSS data using superhorizon perturbation}}, \href{https://doi.org/10.1088/1475-7516/2021/07/035}{\emph{\jcap} {\bfseries 2021} (2021) 035} [\href{https://arxiv.org/abs/2101.11016}{{\ttfamily 2101.11016}}].

\bibitem{Whitford2023}
A.M.~Whitford, C.~Howlett and T.M.~Davis, \emph{Evaluating bulk flow estimators for cosmicflows–4 measurements}, \href{https://doi.org/10.1093/mnras/stad2764}{\emph{Monthly Notices of the Royal Astronomical Society} {\bfseries 526} (2023) 3051–3071}.

\bibitem{KingEllis1973}
A.R.~{King} and G.F.R.~{Ellis}, \emph{{Tilted homogeneous cosmological models}}, \href{https://doi.org/10.1007/BF01646266}{\emph{Communications in Mathematical Physics} {\bfseries 31} (1973) 209}.

\bibitem{Matzner1980}
R.A.~{Matzner}, \emph{{On observations of the cosmic radiation background}}, \href{https://doi.org/10.1086/158397}{\emph{\apj} {\bfseries 241} (1980) 851}.

\bibitem{Ebrahimian2024}
E.~{Ebrahimian}, C.~{Krishnan}, R.~{Mondol} and M.M.~{Sheikh-Jabbari}, \emph{{Towards a realistic dipole cosmology: the dipole {\ensuremath{\Lambda}}CDM model}}, \href{https://doi.org/10.1088/1361-6382/ad550d}{\emph{Classical and Quantum Gravity} {\bfseries 41} (2024) 145007} [\href{https://arxiv.org/abs/2305.16177}{{\ttfamily 2305.16177}}].

\bibitem{Allahyari2023}
A.~Allahyari, E.~Ebrahimian, R.~Mondol and M.M.~Sheikh-Jabbari, \emph{{Big Bang in Dipole Cosmology}},  7, 2023.

\bibitem{Krishnan2022}
C.~Krishnan, R.~Mondol and M.M.~Sheikh-Jabbari, \emph{{A tilt instability in the cosmological principle}}, \href{https://doi.org/10.1140/epjc/s10052-023-12048-y}{\emph{Eur. Phys. J. C} {\bfseries 83} (2023) 874} [\href{https://arxiv.org/abs/2211.08093}{{\ttfamily 2211.08093}}].

\bibitem{Constantin2023}
A.~{Constantin}, T.R.~{Harvey}, S.~{von Hausegger} and A.~{Lukas}, \emph{{Spatially homogeneous universes with late-time anisotropy}}, \href{https://doi.org/10.1088/1361-6382/ad0b36}{\emph{Classical and Quantum Gravity} {\bfseries 40} (2023) 245015} [\href{https://arxiv.org/abs/2212.03234}{{\ttfamily 2212.03234}}].

\bibitem{Cheng2023}
Y.-T.~{Cheng}, T.-C.~{Chang} and A.~{Lidz}, \emph{{Is the Radio Source Dipole from NVSS Consistent with the CMB and $\Lambda$CDM?}}, \href{https://doi.org/10.48550/arXiv.2309.02490}{\emph{arXiv e-prints} (2023) arXiv:2309.02490} [\href{https://arxiv.org/abs/2309.02490}{{\ttfamily 2309.02490}}].

\bibitem{Dalang2021}
C.~Dalang and C.~Bonvin, \emph{{On the kinematic cosmic dipole tension}}, \href{https://doi.org/10.1093/mnras/stac726}{\emph{Mon. Not. Roy. Astron. Soc.} {\bfseries 512} (2022) 3895} [\href{https://arxiv.org/abs/2111.03616}{{\ttfamily 2111.03616}}].

\bibitem{Guandalin2023}
C.~{Guandalin}, J.~{Piat}, C.~{Clarkson} and R.~{Maartens}, \emph{{Theoretical Systematics in Testing the Cosmological Principle with the Kinematic Quasar Dipole}}, \href{https://doi.org/10.3847/1538-4357/acdf46}{\emph{\apj} {\bfseries 953} (2023) 144} [\href{https://arxiv.org/abs/2212.04925}{{\ttfamily 2212.04925}}].

\bibitem{planck2020-LVII}
{\sorthelp{Planck Collaboration IntZZG}}{Planck Collaboration Int. LVII}, \emph{{\textit{Planck} intermediate results. LVII. NPIPE: Joint \Planck\ LFI and HFI data processing}}, \href{https://doi.org/10.1051/0004-6361/202038073}{\emph{\aap} {\bfseries 643} (2020) 42} [\href{https://arxiv.org/abs/2007.04997}{{\ttfamily 2007.04997}}].

\bibitem{gorski2005}
K.M.~{G{\'o}rski}, E.~{Hivon}, A.J.~{Banday}, B.D.~{Wandelt}, F.K.~{Hansen}, M.~{Reinecke} et~al., \emph{{HEALPix: A Framework for High-Resolution Discretization and Fast Analysis of Data Distributed on the Sphere}}, \href{https://doi.org/10.1086/427976}{\emph{\apj} {\bfseries 622} (2005) 759} [\href{https://arxiv.org/abs/astro-ph/0409513}{{\ttfamily astro-ph/0409513}}].

\bibitem{Moss2011}
A.~{Moss}, D.~{Scott}, J.P.~{Zibin} and R.~{Battye}, \emph{{Tilted physics: A cosmologically dipole-modulated sky}}, \href{https://doi.org/10.1103/PhysRevD.84.023014}{\emph{\prd} {\bfseries 84} (2011) 023014} [\href{https://arxiv.org/abs/1011.2990}{{\ttfamily 1011.2990}}].

\bibitem{Bunn1994}
E.~{Bunn}, Y.~{Hoffman} and J.~{Silk}, \emph{{The Effects of Incomplete Sky Coverage on the Analysis of Large Angular Scale Microwave Background Anisotropy}}, \href{https://doi.org/10.1086/173991}{\emph{\apj} {\bfseries 425} (1994) 359}.

\bibitem{Hoffman1991}
Y.~{Hoffman} and E.~{Ribak}, \emph{{Constrained Realizations of Gaussian Fields: A Simple Algorithm}}, \href{https://doi.org/10.1086/186160}{\emph{\apjl} {\bfseries 380} (1991) L5}.

\bibitem{Eriksen2004}
H.K.~{Eriksen}, I.J.~{O'Dwyer}, J.B.~{Jewell}, B.D.~{Wandelt}, D.L.~{Larson}, K.M.~{G{\'o}rski} et~al., \emph{{Power Spectrum Estimation from High-Resolution Maps by Gibbs Sampling}}, \href{https://doi.org/10.1086/425219}{\emph{\apjs} {\bfseries 155} (2004) 227} [\href{https://arxiv.org/abs/astro-ph/0407028}{{\ttfamily astro-ph/0407028}}].

\bibitem{Tegmark1997}
M.~{Tegmark}, \emph{{How to measure CMB power spectra without losing information}}, \href{https://doi.org/10.1103/PhysRevD.55.5895}{\emph{\prd} {\bfseries 55} (1997) 5895} [\href{https://arxiv.org/abs/astro-ph/9611174}{{\ttfamily astro-ph/9611174}}].

\bibitem{Tegmark2001}
M.~Tegmark and A.~de~Oliveira-Costa, \emph{How to measure cmb polarization power spectra without losing information}, \href{https://doi.org/10.1103/physrevd.64.063001}{\emph{Physical Review D} {\bfseries 64} (2001) }.

\bibitem{Wandelt2001}
B.D.~Wandelt, E.~Hivon and K.M.~G\'orski, \emph{Cosmic microwave background anisotropy power spectrum statistics for high precision cosmology}, \href{https://doi.org/10.1103/PhysRevD.64.083003}{\emph{Phys. Rev. D} {\bfseries 64} (2001) 083003} [\href{https://arxiv.org/abs/astro-ph/0008111}{{\ttfamily astro-ph/0008111}}].

\bibitem{Hivon2002}
E.~{Hivon}, K.M.~{G{\'o}rski}, C.B.~{Netterfield}, B.P.~{Crill}, S.~{Prunet} and F.~{Hansen}, \emph{Master of the cosmic microwave background anisotropy power spectrum: A fast method for statistical analysis of large and complex cosmic microwave background data sets}, \href{https://doi.org/10.1086/338126}{\emph{The Astrophysical Journal} {\bfseries 567} (2002) 2–17} [\href{https://arxiv.org/abs/astro-ph/0105302}{{\ttfamily astro-ph/0105302}}].

\bibitem{Szapudi2001}
I.~{Szapudi}, S.~{Prunet}, D.~{Pogosyan}, A.S.~{Szalay} and J.R.~{Bond}, \emph{{Fast Cosmic Microwave Background Analyses via Correlation Functions}}, \href{https://doi.org/10.1086/319105}{\emph{\apjl} {\bfseries 548} (2001) L115}.

\bibitem{Chon2004}
G.~{Chon}, A.~{Challinor}, S.~{Prunet}, E.~{Hivon} and I.~{Szapudi}, \emph{{Fast estimation of polarization power spectra using correlation functions}}, \href{https://doi.org/10.1111/j.1365-2966.2004.07737.x}{\emph{\mnras} {\bfseries 350} (2004) 914} [\href{https://arxiv.org/abs/astro-ph/0303414}{{\ttfamily astro-ph/0303414}}].

\bibitem{anesthetic}
W.~Handley, \emph{anesthetic: nested sampling visualisation}, \href{https://doi.org/10.21105/joss.01414}{\emph{The Journal of Open Source Software} {\bfseries 4} (2019) 1414}.

\bibitem{Dam2022}
L.~{Dam}, G.F.~{Lewis} and B.J.~{Brewer}, \emph{{Testing the Cosmological Principle with CatWISE Quasars: A Bayesian Analysis of the Number-Count Dipole}}, \href{https://doi.org/10.48550/arXiv.2212.07733}{\emph{arXiv e-prints} (2022) arXiv:2212.07733} [\href{https://arxiv.org/abs/2212.07733}{{\ttfamily 2212.07733}}].

\bibitem{Singal2024}
A.K.~{Singal}, \emph{{Resolution of the incongruency of dipole asymmetries within various large radio surveys - implications for the Cosmological Principle}}, \href{https://doi.org/10.1093/mnras/stae414}{\emph{\mnras} {\bfseries 528} (2024) 5679} [\href{https://arxiv.org/abs/2312.12785}{{\ttfamily 2312.12785}}].

\bibitem{Mittal2023}
V.~{Mittal}, O.T.~{Oayda} and G.F.~{Lewis}, \emph{{The cosmic dipole in the Quaia sample of quasars: a Bayesian analysis}}, \href{https://doi.org/10.1093/mnras/stad3706}{\emph{\mnras} {\bfseries 527} (2024) 8497} [\href{https://arxiv.org/abs/2311.14938}{{\ttfamily 2311.14938}}].

\bibitem{Quaia2023}
K.~{Storey-Fisher}, D.W.~{Hogg}, H.-W.~{Rix}, A.-C.~{Eilers}, G.~{Fabbian}, M.~{Blanton} et~al., \emph{{Quaia, the Gaia-unWISE Quasar Catalog: An All-Sky Spectroscopic Quasar Sample}}, \href{https://doi.org/10.48550/arXiv.2306.17749}{\emph{arXiv e-prints} (2023) arXiv:2306.17749} [\href{https://arxiv.org/abs/2306.17749}{{\ttfamily 2306.17749}}].

\end{thebibliography}\endgroup

\appendix
\section{Masking and multipole mixing}

\label{sec:appendidx}
Here we investigate how masking causes leakage from other multipoles into the dipole, making dipole estimation challenging. We follow the derivation from Ref.~\cite{Wandelt2001}. Decomposition into different spherical harmonic modes is a useful technique for analysing the temperature anisotropies in the CMB, or indeed any all-sky data set. A spin-zero field can be decomposed into spherical harmonic coefficients as
\begin{equation}
T(\theta, \phi)  = \sum_{\ell,m} a_{\ell m} Y_{\ell m} (\theta, \phi).
\end{equation}
We assume that the sky has been observed only over a region $\mathcal{O}$ through the application of a mask.

The coefficients in the spherical harmonic expansion recovered from such an incomplete observing region are
\begin{equation}
\hat{a}_{l^{\prime} m^{\prime}} =\int_{\mathcal{O}} d \Omega Y_{l^{\prime} m^{\prime}}^*(\theta, \phi) T(\theta, \phi) =\sum_{l m} a_{l m} \int_{\mathcal{O}} d \Omega Y_{l^{\prime} m^{\prime}}^*(\theta, \phi) Y_{l m}(\theta, \phi).
\end{equation}
The notation $\int_{\mathcal{O}}$ denotes integration over the observed region. Note that the usual orthogonality property of the $Y_{\ell m}(\theta, \phi)$ does not hold any longer because we are not integrating over all solid angles. This becomes clearer if we define the geometric coupling matrix
\begin{equation}
M_{l^{\prime} m^{\prime} l m} \equiv \int_{\mathcal{O}} d \Omega Y_{l^{\prime} m^{\prime}}^*(\theta, \phi) Y_{l m}(\theta, \phi).
\end{equation}
This coupling matrix depends solely on the shape of the mask. For the mask used in this study, we have calculated and shown the matrix in \cref{fig:coupling_matrix}. 

\end{document}